\documentclass[a4paper]{jpconf}

\usepackage{graphicx}
\newcommand{\de}{{\rm d}}
\usepackage{bm}

\begin{document}
\title{Azimuthal asymmetries for hadron distributions inside a jet in hadronic collisions}

\author{Umberto D'Alesio$^{1, 2}$, Francesco Murgia$^2$ and Cristian Pisano$^{1,2, }$\footnote[99]{Speaker. Talk presented at the 19th International Spin Physics Symposium (SPIN2010),
September 27 - October 2, 2010, J\"ulich (Germany).}}

\address{$^1$ Dipartimento di Fisica, Universit\`a di Cagliari, Cittadella Universitaria, \\ I-09042 Monserrato (CA), Italy}

\address{$^2$ Istituto Nazionale di Fisica Nucleare, Sezione di Cagliari, C.P.\ 170,\\ I-09042 Monserrato (CA), Italy}

\ead{umberto.dalesio@ca.infn.it, francesco.murgia@ca.infn.it, cristian.pisano@ca.infn.it}

\begin{abstract}
Using a generalized parton model approach including spin and intrinsic parton motion effects, and assuming the validity of factorization for large $p_T$ jet production in hadronic collisions, we study the azimuthal
distribution around the jet axis of leading pions, produced in the jet fragmentation process. We identify the observable
leading-twist azimuthal asymmetries for the unpolarized
and single-polarized case related to both quark and gluon-originated jets.
We account for all physically allowed combinations of the transverse momentum  dependent (TMD) parton distribution and fragmentation functions, with
special attention to the Sivers, Boer-Mulders, and transversity
quark distributions, and to the Collins fragmentation function for quarks
(and to the analogous functions for gluon partons). 
\end{abstract}


\section{\label{intro} Introduction}

Transverse single-spin and azimuthal asymmetries in high-energy hadronic
reactions have raised a lot of interest in the last years
(see e.g.~Refs.~\cite{D'Alesio:2007jt,Barone:2010ef} and references therein).
In particular, the huge spin asymmetries measured in the inclusive 
forward production of pions in high-energy $pp$ collisions, at moderately large transverse momentum, cannot be explained in the realm of
 leading-twist (LT) perturbative QCD (pQCD),
based on the usual collinear factorization theorems. 

Out of the theoretical approaches proposed 
 in order to account for these measurements, in the following we 
will  adopt the so-called transverse momentum dependent
(TMD) formalism, which takes into
account spin and intrinsic parton motion effects assuming a pQCD 
factorization scheme. Single-spin and azimuthal asymmetries are generated
by TMD polarized partonic distribution and fragmentation functions, among which
the most relevant from a phenomenological point of view are the
Sivers distribution~\cite{Sivers:1989cc,Sivers:1990fh}
and, for transversely polarized quarks,
the Boer-Mulders distribution~\cite{Boer:1997nt}
and the Collins fragmentation function~\cite{Collins:1992kk}
(similar functions can be defined for linearly polarized gluons,
see e.g.~Ref.~\cite{Anselmino:2005sh}).

Along the lines of \cite{D'Alesio:2010am}, we consider the process 
$p^{(\uparrow)}p\to{\rm jet}+\pi+X$,  presently under active investigation at RHIC, where one observes a large $p_T$ jet
and looks for the azimuthal distribution of leading pions inside the jet.
A very preliminary version of this study was first presented
in Ref.~\cite{dalesio:2007ec}.
A similar analysis was performed in Ref.~\cite{Yuan:2007nd}, which
however considered intrinsic parton motion only in the fragmentation
process, drastically reducing the possible contributions to the
asymmetry. Indeed, in that case, only the Collins effect for
quarks is at work. In fact, Ref.~\cite{Yuan:2007nd} aimed at
studying only the Collins fragmentation function (FF),
which should be universal, in
a more simplified theoretical scheme
for which factorization has been proven.
Our approach is different in some respects. It is more general
and has in principle a richer structure in the observable
azimuthal asymmetries, since intrinsic motion is taken into
account in the initial hadrons also.
However, since factorization has not been proven in this case, but
is rather taken as a reasonable phenomenological assumption,
the validity of the scheme and the universality of the TMD
distributions involved require an even more severe scrutiny
by comparison with experimental results.
On the other hand, at the present theoretical and
experimental stage, we believe that combined phenomenological
tests of different approaches are required to clarify the
validity of factorization and, related to this, the relevance
of possible universality-breaking terms for the TMD distributions.

The plan of this contribution  is as follows. 
In Sec.~\ref{sec-formalism} we
will briefly summarize the TMD generalized parton model approach
and give the expression of the polarized cross section for the
process of interest.
In Sec.~\ref{sec-results} we will present phenomenological
results for  azimuthal asymmetries discussed in the
kinematical configuration of the RHIC experiments. Sec.~4 contains final remarks and conclusions.

\section{\label{sec-formalism} Formalism}

We denote with  $A$ and  $B$ two spin $1/2$ hadrons
(typically, two protons), with hadron $B$ unpolarized and hadron $A$ in a pure transverse spin state described by the four-vector $S_A$.
Within a generalized TMD parton model approach, 
the invariant differential cross section for the process $A(S_A)\, B\to {\rm jet}+\pi+X$
can be written, at LT in the soft TMD functions, as
\begin{eqnarray}
\frac{E_{\rm j}\,\de\sigma}
{\de^3\bm{p}_{\rm j}\,\de z\,\de^2\bm{k}_{\perp\pi}} &=&
\sum_{a,b,c,d,\{\lambda\}} \int \frac{\de x_a \de x_b}{16\pi^2 x_a x_b s}\,
\de^2\bm{k}_{\perp a}\de^2\bm{k}_{\perp b}\,
\rho^{a/A,S_A}_{\lambda^{}_a\lambda^{\prime}_a}\hat{f}_{a/A,S_A}(x_a,\bm{k}_{\perp a})
\nonumber\\
&\times& \!\!\! \rho^{b/B}_{\lambda^{}_b\lambda^{\prime}_b}\hat{f}_{b/B}(x_b,\bm{k}_{\perp b}) \,\hat{M}_{\lambda^{}_c,\lambda^{}_d;\lambda^{}_a,\lambda^{}_b}
\hat{M}^*_{\lambda^{\prime}_c,\lambda^{}_d;\lambda^{\prime}_a,\lambda^{\prime}_b}
\delta(\hat{s}+\hat{t}+\hat{u})\hat{D}^\pi_{\lambda^{}_c,\lambda^{\prime}_c}(z,\bm{k}_{\perp\pi})\,, 
\label{ds-pi-jet}
\end{eqnarray}
where $E_{\rm j}$ and $\bm{p}_{\rm j}$ are respectively the energy
and three-momentum of the observed jet. We sum over all allowed partonic processes contributing to the physical process observed, and
$\{\lambda\}$ stays for a sum over all partonic helicities. $x_{a,b}$ and $\bm{k}_{\perp a,b}$ are respectively the initial parton light-cone
momentum fractions and intrinsic transverse momenta. Analogously, $z$ and $\bm{k}_{\perp\pi}$ are
the light-cone momentum fraction and the transverse momentum of the observed pion inside the jet w.r.t.\  the jet (parton $c$) direction of motion. 

All information on the polarization state of the initial parton $a$ is contained in  $\rho^{a/A,S_A}_{\lambda^{}_a\lambda^{\prime}_a}\hat{f}_{a/A,S_A}(x_a,\bm{k}_{\perp a})$, which depends
in turn on the (experimentally fixed) parent hadron $A$ polarization state and on the
soft, nonperturbative dynamics encoded in the eight leading-twist polarized and transverse
momentum dependent parton distribution functions.
 $\rho^{a/A,S_A}_{\lambda^{}_a\lambda^{\prime}_a}$ is the helicity
density matrix of parton $a$.
Analogously, the polarization state of parton $b$ inside the unpolarized
hadron $B$ is encoded into
$\rho^{b/B}_{\lambda^{}_b\lambda^{\prime}_b}\hat{f}_{b/B}(x_b,\bm{k}_{\perp b})$. 
The $\hat{M}_{\lambda^{}_c,\lambda^{}_d;\lambda^{}_a,\lambda^{}_b}$'s are the
pQCD leading-order (LO) helicity scattering amplitudes for the hard partonic process $ab\to cd$. 
The $\hat{D}^\pi_{\lambda^{}_c,\lambda^{\prime}_c}(z,\bm{k}_{\perp\pi})$'s are the soft
leading-twist TMD fragmentation functions describing the fragmentation process of the
scattered (polarized) parton $c$ into the final leading pion inside the jet.
More details  can be found in Ref.~\cite{D'Alesio:2010am}.

We work in the $AB$ hadronic c.m.~frame, with hadron $A$ moving
along the $+\hat{\bm{Z}}_{\rm cm}$ direction, and define $(XZ)_{\rm cm}$ as the
production plane containing the colliding beams and the observed jet,
with $(\bm{p}_{\rm j})_{X_{\rm cm}}>0$. In this frame $S_A = (0, \cos\phi_{S_A},\sin\phi_{S_A},0) $ and $p_{\rm j} = 
p_{{\rm j}\,T}(\cosh \eta_{\rm j},1,0,\sinh \eta_{\rm j})$,
where $\eta_{\rm j} = -\log[\tan(\theta_{\rm j}/2)]$ is the jet (pseudo)rapidity.

The calculation is performed by summing explicitly
over all helicity indexes and inserting the appropriate expressions for the
helicity density matrices of partons $a$, $b$ and for the polarized distribution and fragmentation functions.
After factorizing explicitly all azimuthal dependences,
including those coming from the hard-scattering helicity amplitudes, collecting
them and using symmetry properties under $\bm{k}_{\perp a,b}\to - \bm{k}_{\perp a,b}$ \cite{D'Alesio:2010am},
one gets the final expression for the single transverse polarized cross 
section. This will have
the following general structure:
\begin{eqnarray}
2{\rm d}\sigma(\phi_{S_A},\phi_\pi^H) \!\!\! &\sim& \!\!\! {\rm d}\sigma_0
+{\rm d}\Delta\sigma_0\sin\phi_{S_A}+
{\rm d}\sigma_1\cos\phi_\pi^H+ {\rm d}\sigma_2\cos2\phi_\pi^H+
{\rm d}\Delta\sigma_{1}^{-}\sin(\phi_{S_A}-\phi_\pi^H)
\nonumber\\
\!\!\!&+& \!\!\!{\rm d}\Delta\sigma_{1}^{+}\sin(\phi_{S_A}+\phi_\pi^H)
+{\rm d}\Delta\sigma_{2}^{-}\sin(\phi_{S_A}-2\phi_\pi^H)+
{\rm d}\Delta\sigma_{2}^{+}\sin(\phi_{S_A}+2\phi_\pi^H)\,,
\label{d-sig-phi-SA}
\end{eqnarray}
where $\phi_\pi^H$ is the azimuthal
angle of the pion three-momentum around the jet direction of motion, as measured in the fragmenting parton helicity frame. The latter frame is related to the hadronic c.m.~frame by a simple rotation
by $\theta_{\rm j}$ around $\hat{\bm{Y}}_{\rm cm}\equiv \hat{\bm{y}}_{\rm j}$ \cite{D'Alesio:2010am}.

In terms of the polarized cross section in  Eq.~(\ref{d-sig-phi-SA}), we can define
average values of appropriate circular functions of $\phi_{S_A}$ and $\phi_\pi^H$,
in order to single out the different contributions of interest:
\begin{equation}
\langle\,W(\phi_{S_A},\phi_\pi^H)\,\rangle(\bm{p}_{\rm j},z,k_{\perp\pi})=
\frac{\int{\rm d}\phi_{S_A}{\rm d}\phi_\pi^H\,
W(\phi_{S_A},\phi_\pi^H)\,{\rm d}\sigma(\phi_{S_A},\phi_\pi^H)}
{\int{\rm d}\phi_{S_A}{\rm d}\phi_\pi^H{\rm\, d}\sigma(\phi_{S_A},\phi_\pi^H)}\,.
\label{average}
\end{equation}
Alternatively, for the single spin asymmetry we can, in close analogy with the case of semi-inclusive
deeply inelastic scattering (SIDIS), define appropriate azimuthal moments,
\begin{eqnarray}
A_N^{W(\phi_{S_A},\phi_\pi^H)}(\bm{p}_{\rm j},z,k_{\perp\pi})
\!\!&\equiv&\!\!
2\langle\,W(\phi_{S_A},\phi_\pi^H)\,\rangle(\bm{p}_{\rm j},z,k_{\perp\pi})\nonumber\\
\!\!&=&\!\!
2\,\frac{\int{\rm d}\phi_{S_A}{\rm d}\phi_\pi^H\,
W(\phi_{S_A},\phi_\pi^H)\,[{\rm d}\sigma(\phi_{S_A},\phi_\pi^H)-
{\rm d}\sigma(\phi_{S_A}+\pi,\phi_\pi^H)]}
{\int{\rm d}\phi_{S_A}{\rm d}\phi_\pi^H\,
[{\rm d}\sigma(\phi_{S_A},\phi_\pi^H)+
{\rm d}\sigma(\phi_{S_A}+\pi,\phi_\pi^H)]}\,,
\label{gen-mom}
\end{eqnarray}
where $W(\phi_{S_A},\phi_\pi^H)$ is again some appropriate circular function of $\phi_{S_A}$
and $\phi_\pi^H$.

\section{\label{sec-results} Phenomenology}

In this section we present and discuss some phenomenological implications
of our approach for
the unpolarized and single-transverse polarized cases in kinematical
configurations accessible at RHIC by the STAR and PHENIX experiments.
We consider both central ($\eta_{\rm j}=0$) and forward
($\eta_{\rm j}=3.3$) (pseudo)rapidity configurations
at a c.m.~energy $\sqrt{s} =$ 200 GeV (different c.m.~energies, namely $\sqrt{s}=62.4$ and 500 GeV, are also studied in \cite{D'Alesio:2010am}),
aiming at a check of the potentiality of the approach in
disentangling among different quark and gluon originating effects.

We will first consider, for $\pi^+$ production only,
a scenario in which the effects of
all TMD functions are over-maximized. By this we mean that all TMD
functions are maximized in size by imposing natural positivity bounds
(and the Soffer bound for transversity~\cite{Soffer:1994ww,Bacchetta:1999kz});
moreover, the relative signs of
all active partonic contributions are chosen so that they
sum up additively. In this way we set
an upper bound on the absolute value of any of the effects playing
a potential role in the azimuthal asymmetries.
Therefore, all effects that are negligible or even
marginal in this scenario may be directly discarded in subsequent
refined phenomenological analyses.

As a second step in our study we consider, for both neutral and charged pions,
only the surviving effects, involving TMD functions for which parameterizations
are available from independent fits to other spin and azimuthal
asymmetries data in SIDIS, Drell-Yan, and $e^+e^-$ processes.

For numerical calculations all TMD distribution and
fragmentation functions will be taken in the simplified form where
the functional dependences on
the parton light-cone momentum fraction and on transverse
motion are completely factorized, assuming a Gaussian-like flavour-independent
shape for the transverse momentum component.
Concerning the parameterizations of the transversity and quark 
Sivers distributions,
 and
of the Collins functions, we will consider two sets:
SIDIS~1 \cite{Anselmino:2005ea, Anselmino:2007fs} and  
SIDIS~2 \cite{Anselmino:2008sga,Anselmino:2008jk}. 
Notice that the almost unknown gluon Sivers function was tentatively
taken positive and saturated to an updated version of the bound obtained
in Ref.~\cite{Anselmino:2006yq} by considering
PHENIX data for the $\pi^0$ transverse SSA at mid-rapidity production in
polarized $pp$ collisions at RHIC~\cite{Adler:2005in}.
Furthermore, for the usual collinear distributions, we  adopt the LO 
unpolarized set GRV98~\cite{Gluck:1998xa} and (for the Soffer bound) the 
corresponding longitudinally
polarized set GRSV2000~\cite{Gluck:2000dy}. For fragmentation functions, we will adopt two well-known LO sets
among those available in the literature, the set by Kretzer~\cite{Kretzer:2000yf} and the DSS one~\cite{deFlorian:2007aj}.
Our choice is dictated by the subsequent use of the two available 
parametrization sets for the Sivers and Collins functions
in our scheme, that have been extracted in the past years by adopting these sets of FFs.

Since the range of the jet transverse momentum
(the hard scale) covered is significant, we  take into account proper evolution
with scale. Concerning transversity, in the maximized scenario we will fix
it at the initial scale by saturating the Soffer bound and then letting it
evolve. On the other hand, the transverse momentum component of all TMD functions
is kept fixed with no evolution with scale.
Notice that at this stage evolution properties of the full TMD functions are not known.

In all cases considered, since we are interested in azimuthal asymmetries for
leading particles inside the jet, we will present results obtained
integrating the light-cone momentum fraction of the observed hadron, $z$,
in the range $z\geq 0.3$.

\subsection{\label{sec-results-unp} Azimuthal asymmetries in  $p p\to {\rm jet}+\pi+X$\\}

The symmetric part ${\rm d}\sigma_0$  in Eq.~(\ref{d-sig-phi-SA}) gets 
contributions by the usual unpolarized term,
already present in the collinear approach, and by an additional term
involving a Boer-Mulders$\,\otimes\,$Boer-Mulders convolution for the initial
quarks (or the analogous terms involving linearly polarized gluons);
however even  in the maximized
scenario this last contribution is always negligible in all the kinematical
configurations considered,  hence we will not discuss it anymore in the 
sequel.

\begin{figure*}[t]
\begin{center}
 \includegraphics[angle=0,width=0.4\textwidth]{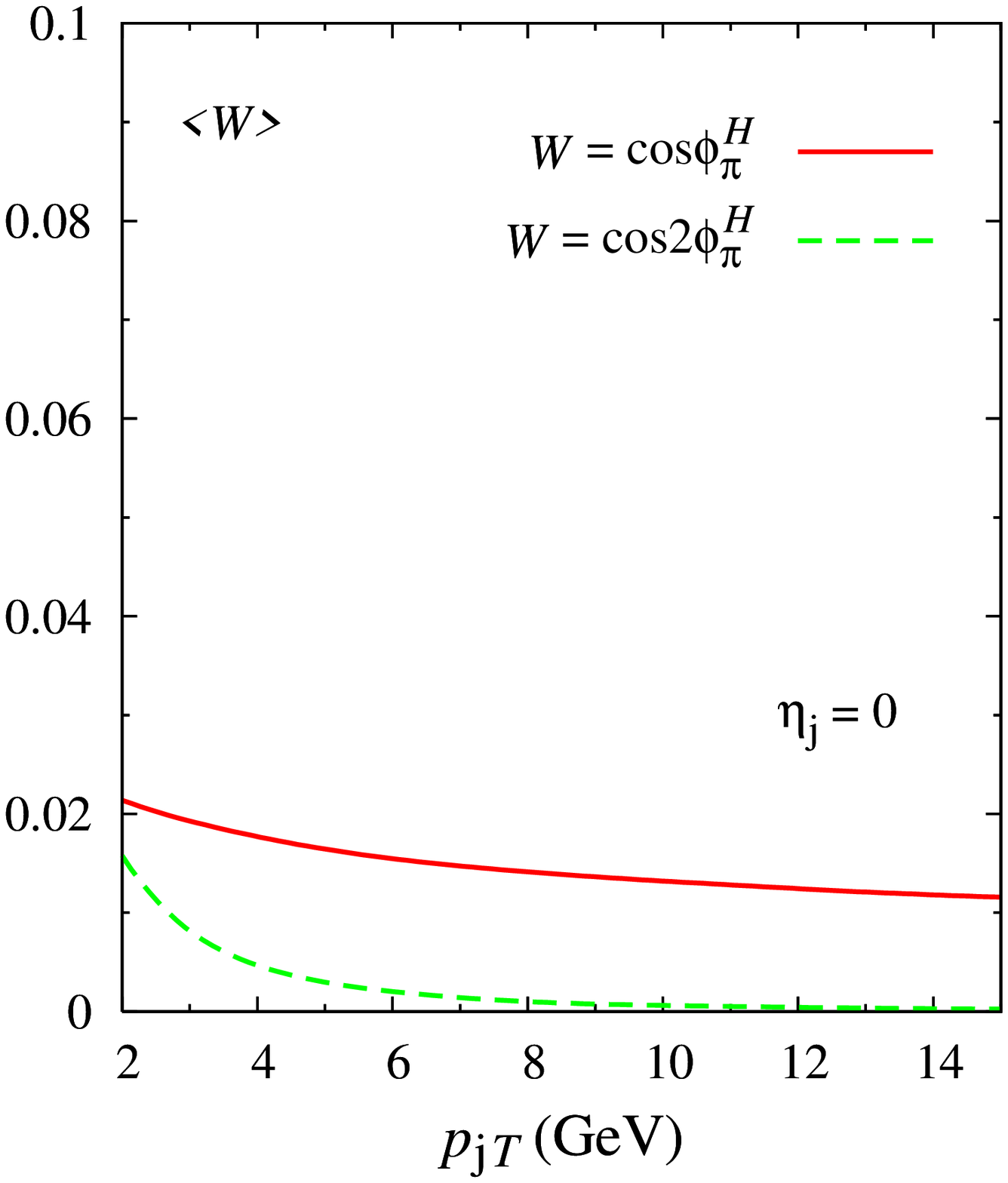}
 \includegraphics[angle=0,width=0.4\textwidth]{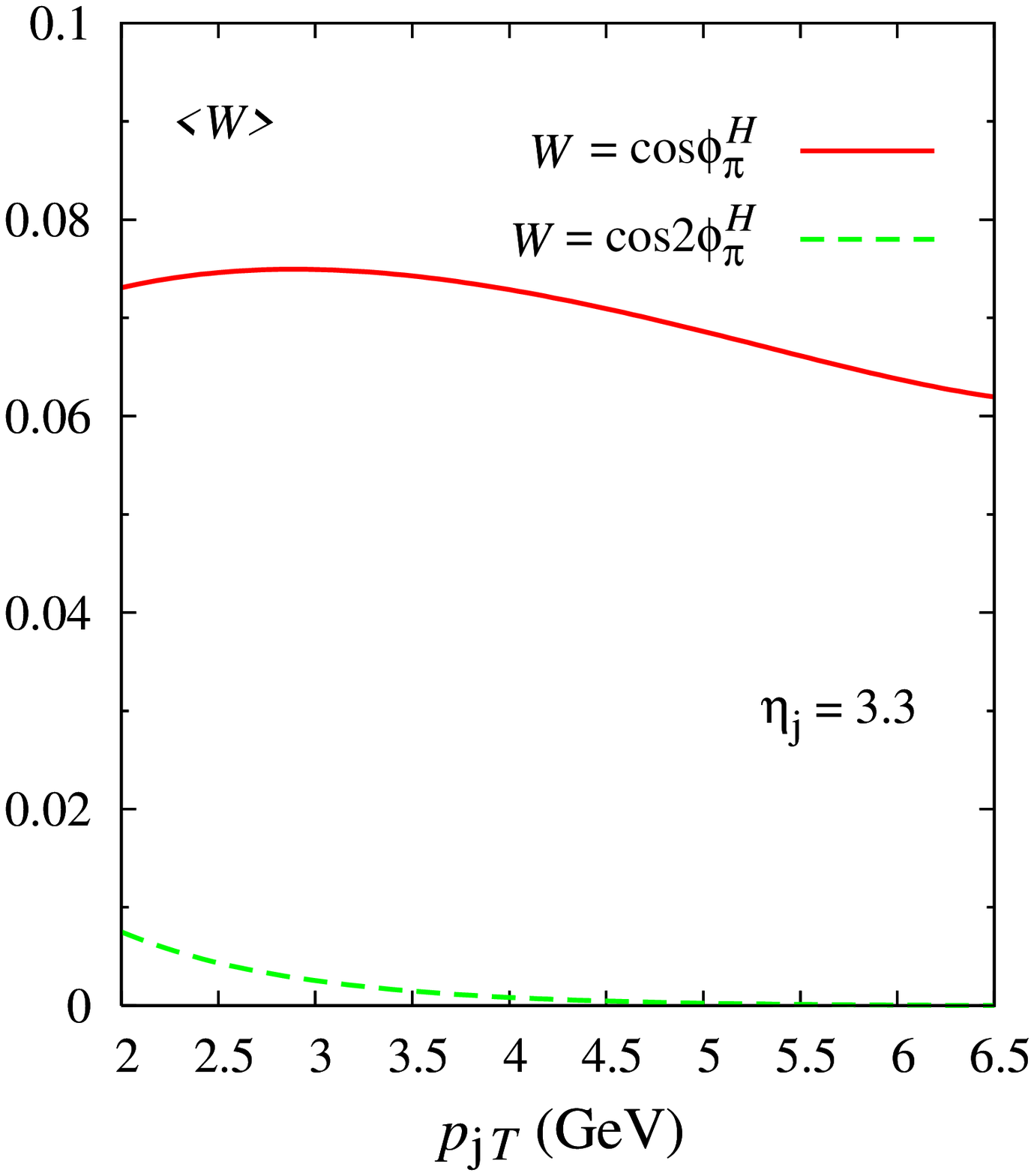}
 \caption{Maximized  quark-originated ($\cos\phi_\pi^H$) and
 gluon-originated ($\cos2\phi_\pi^H$) asymmetries
for the unpolarized $pp\to {\rm jet}+\pi^+ + X$ process
at $\sqrt{s}=200$ GeV in two different rapidity regions, adopting
the Kretzer FF set.
 \label{asy-unp-200} }
\end{center}
\end{figure*}

In Fig.~\ref{asy-unp-200} we show the
maximized  $\langle \cos\phi_\pi^H \rangle $ and  
$\langle\cos2\phi_\pi^H\rangle$
asymmetries
for $\pi^+$ production in the central (left panel)
and forward (right panel) rapidity regions as a function of $p_{{\rm j}\,T}$,
from $p_{{\rm j}\,T}=2$ GeV up to the maximum allowed value, adopting
the Kretzer FF set.  Similar results are obtained using the
DSS set. The $\cos\phi_\pi^H$ asymmetry  is generated by the
quark Boer-Mulders$\,\otimes\,$Collins  convolution term,
involving a transversely polarized quark and an unpolarized hadron
both in the initial state and in the fragmentation process. 
 The $\cos2\phi_\pi^H$ asymmetry is related to the term involving
linearly polarized gluons and unpolarized hadrons both in the
initial state and in the fragmentation process, that is the
convolution of a Boer-Mulders-like gluon distribution with
a Collins-like gluon FF. Even the maximized contribution is
practically negligible in the kinematical configurations considered.

\subsection{\label{sec-results-an-siv} The Sivers asymmetry $A_N^{\sin\phi_{S_A}}$ in $p^\uparrow p\to{\rm jet}+\pi+X$\\}

\begin{figure*}[b]
\begin{center}
 \includegraphics[angle=0,width=0.4\textwidth]{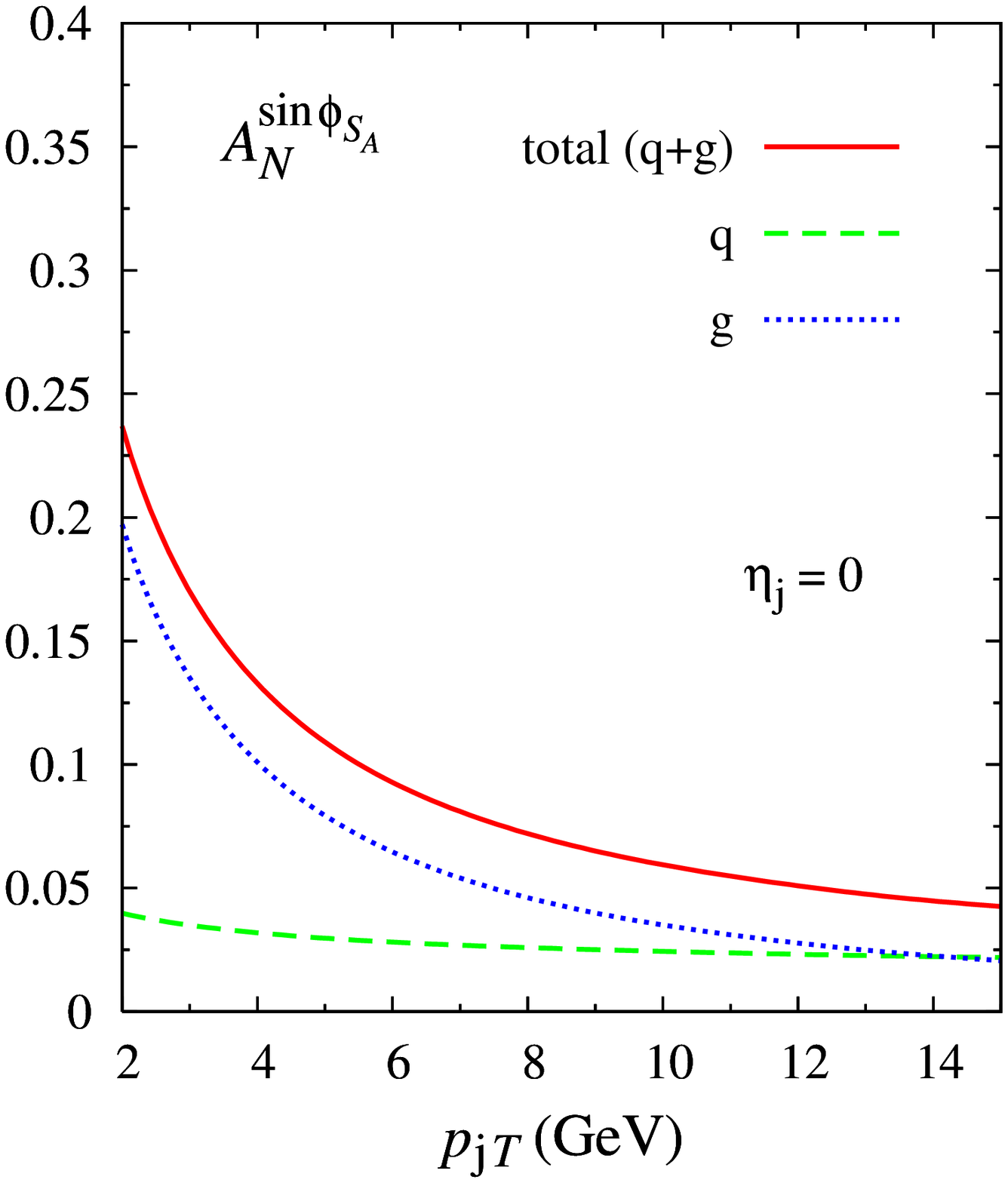}
 \includegraphics[angle=0,width=0.4\textwidth]{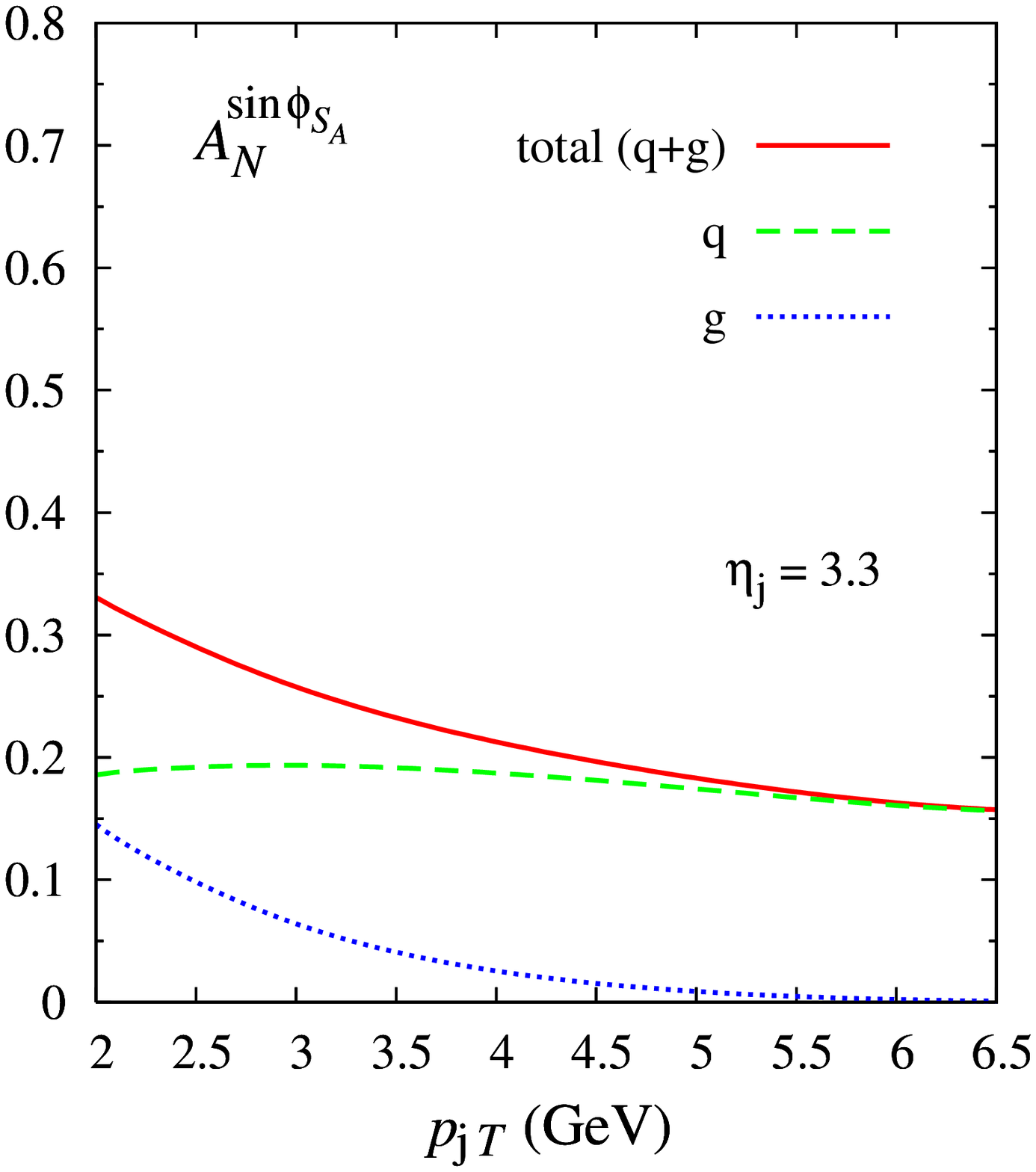}
 \caption{Maximized  total,
quark-originated and
gluon-originated Sivers asymmetries
for the $p^\uparrow p\to {\rm jet}+\pi^+ + X$ process,
at $\sqrt{s}=200$ GeV  in two different  rapidity regions, adopting
the Kretzer FF set.
 \label{asy-an-siv-max200} }
\end{center}
\end{figure*}

In Fig.~\ref{asy-an-siv-max200} we show the total observable Sivers
asymmetry, and the corresponding quark and gluon contributions
 for $\pi^+$ production,
in the maximized scenario and adopting the Kretzer FF
 set, as a function of $p_{{\rm j}\,T}$ in the
central (left panel) and forward (right panel) rapidity regions.
The maximized potential Sivers asymmetry can be very large in both
cases. In the central rapidity region, the asymmetry is dominated by
the gluon contribution at the lowest $p_{{\rm j}\,T}$ range while
gets comparable quark and gluon contributions in the large
$p_{{\rm j}\,T}$ range. A large Sivers asymmetry around
$p_{{\rm j}\,T}=4\div6$ GeV could then be a clear indication for
a sizable gluon contribution. 
In the forward rapidity region, on the contrary, the quark and
gluon contributions are comparable at low $p_{{\rm j}\,T}$ values,
while the maximized asymmetry is dominated by the quark contribution
for $p_{{\rm j}\,T}\ge 4$ GeV. Therefore, a large Sivers
asymmetry in this kinematical range could be ascribed unambiguosly
to the quark Sivers effect.

\begin{figure*}[t]
\begin{center}
 \includegraphics[angle=0,width=0.34\textwidth]{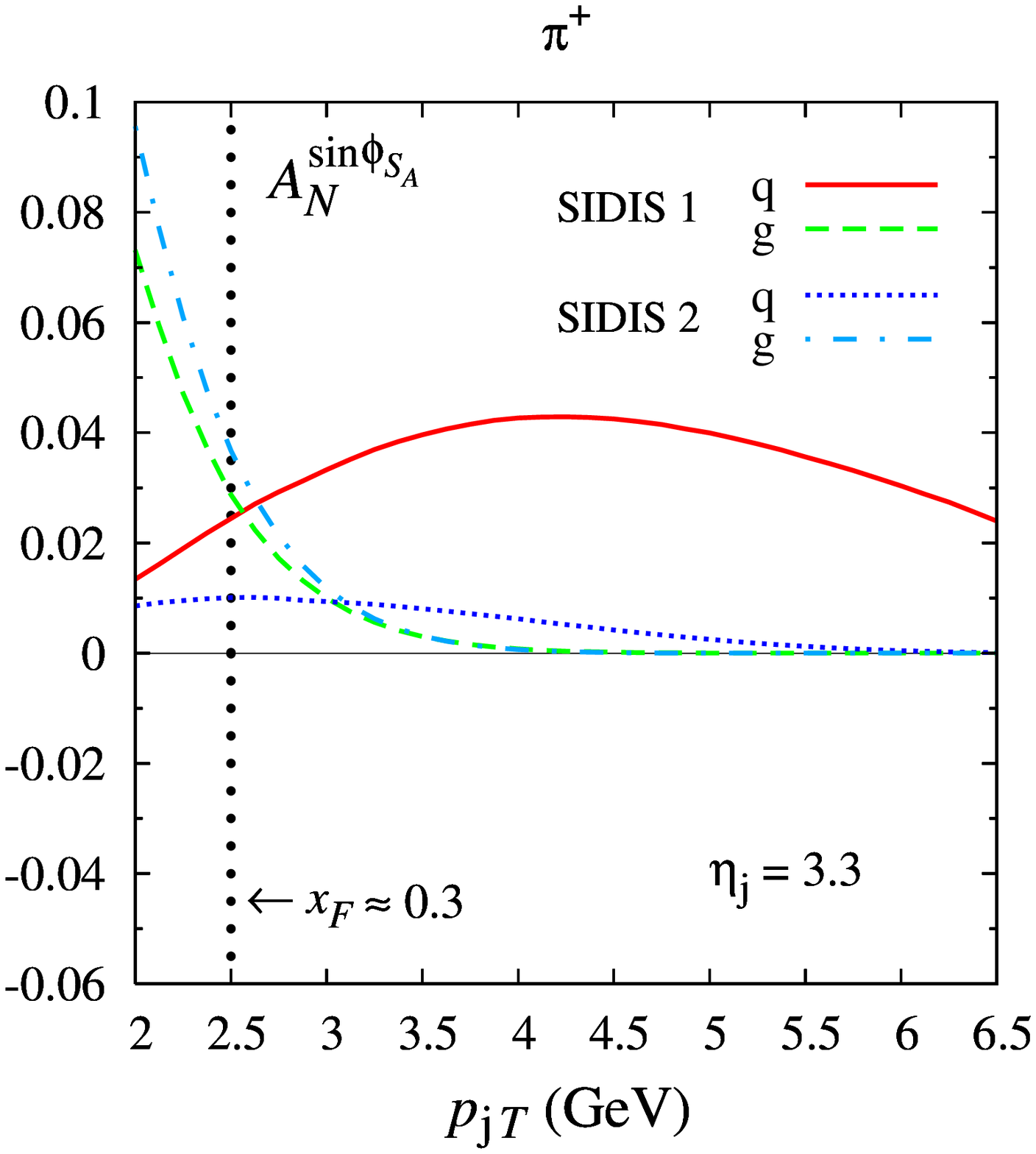}
 \hspace*{-20pt}
 \includegraphics[angle=0,width=0.34\textwidth]{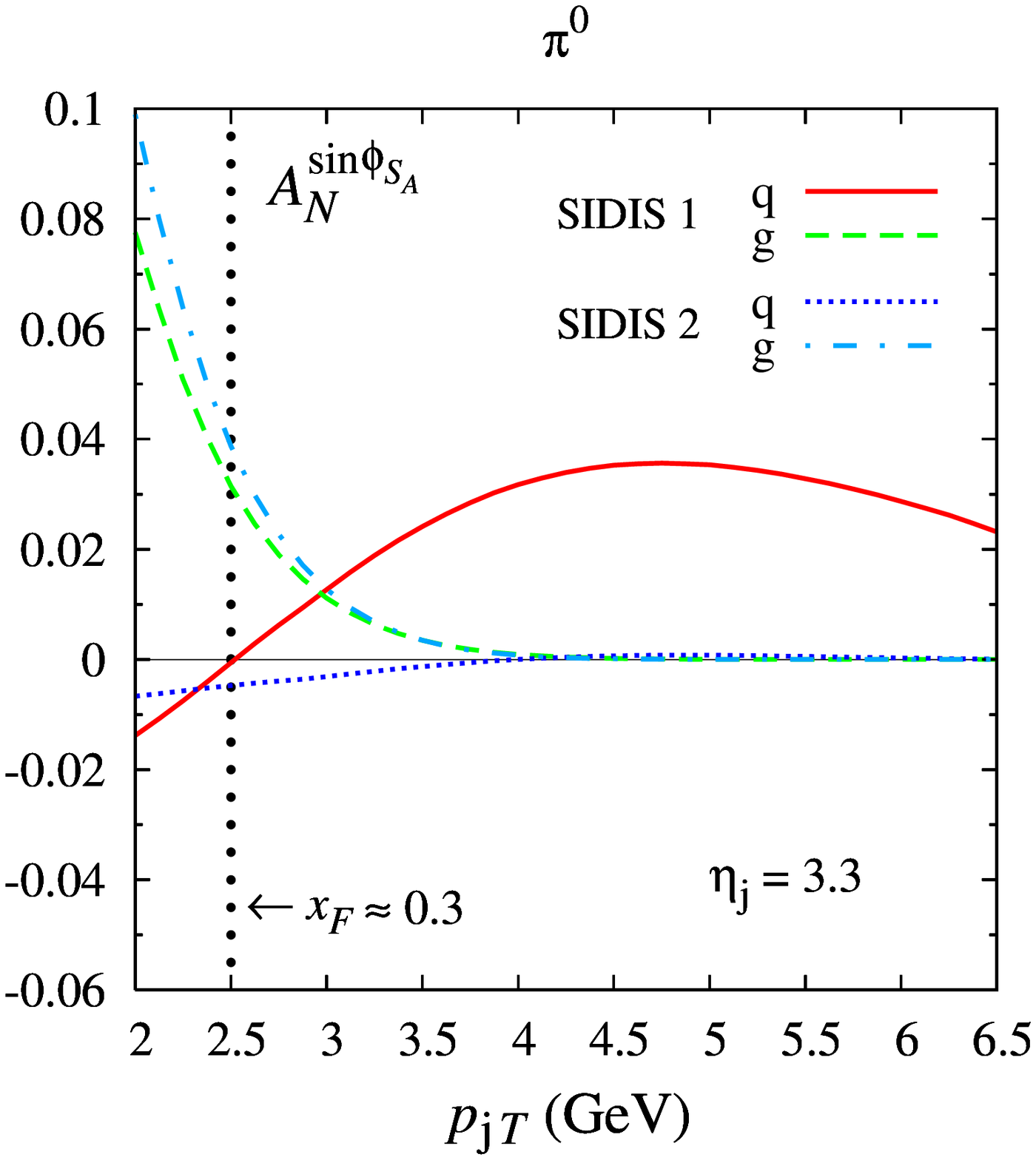}
 \hspace*{-20pt}
 \includegraphics[angle=0,width=0.34\textwidth]{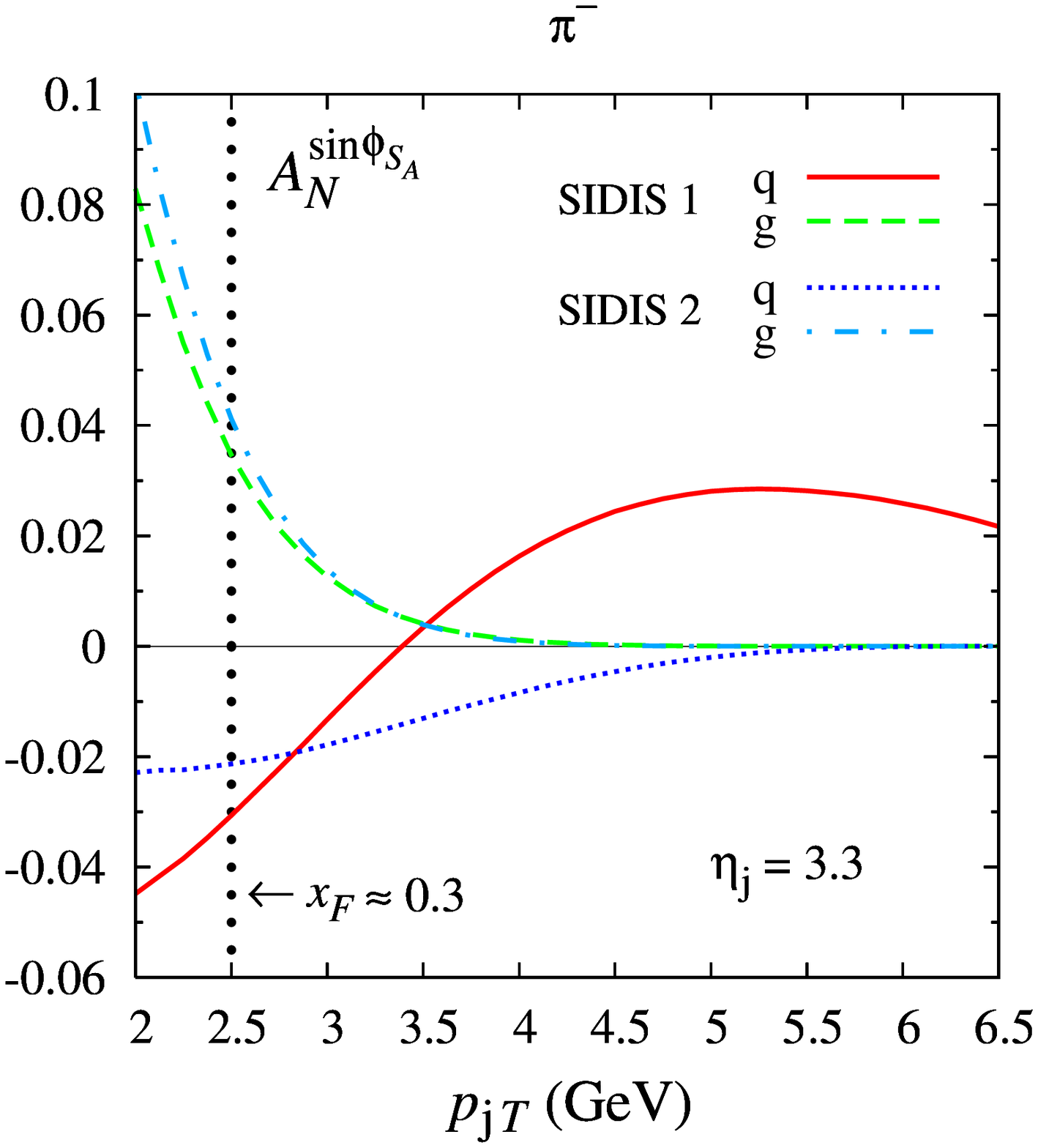}
 \caption{The estimated quark and gluon
 contributions to the Sivers asymmetry
 for the $p^\uparrow p\to {\rm jet}+\pi + X$ process,
 obtained adopting the parametrization sets
 SIDIS~1 and
 SIDIS~2,
at forward rapidity and $\sqrt{s}=200$ GeV.
The dotted black vertical line delimits the region $x_F\approx 0.3$.
 \label{asy-an-siv-par200} }
\end{center}
\end{figure*}

In Fig.~\ref{asy-an-siv-par200} we show,
for both neutral and charged pions, the quark and gluon
contributions to the Sivers asymmetry,
obtained adopting  the parametrization sets SIDIS~1
and SIDIS~2, and the updated version
of the bound on the gluon Sivers asymmetry derived in Ref.~\cite{Anselmino:2006yq}, in the forward rapidity region, as a function of
$p_{{\rm j}\,T}$. The dotted black vertical line delimits the region 
$x_F \approx 0.3$, with $x_F=2 p_{{\rm j}\,L}/\sqrt{s}$,
beyond which the SIDIS parameterizations for the quark Sivers
distribution are extrapolated outside the $x$
region covered by SIDIS data and are therefore plagued by large uncertainties.
This reflects on the fact that below this limit the two sets give comparable
results, while above it they differ remarkably. 
Therefore, a measurement of this asymmetry might help in clarifying
the behaviour of the quark Sivers distribution in the large $x$ region,
which plays a fundamental role for forward pion production at RHIC, and 
is not covered by present SIDIS data from HERMES and COMPASS experiments. 

\subsection{\label{sec-results-an-col}
The Collins(-like) $A_N^{\sin(\phi_{S_A}\mp\phi_\pi^H)}$
 ($A_N^{\sin(\phi_{S_A}\mp2\phi_\pi^H)}$) asymmetries in $p^\uparrow p\to{\rm jet}+\pi+X$\\}

The quark generated asymmetry
$A_N^{\sin(\phi_{S_A}+\phi_\pi^H)}$ comes from two distinct
contributions:
one involving the convolution between the term of
the TMD transversity distribution suppressed in the collinear
configuration 
and the Collins function; another term
involving the convolution of the Sivers and Boer-Mulders
distributions for the initial quarks with the Collins function
for the final quark [an analogous term appears also
in the $A_N^{\sin(\phi_{S_A}-\phi_\pi^H)}$
asymmetry]. We have explicitly checked that for the kinematical 
configurations under study both these
contributions are always negligible already in the maximized scenario.
Therefore we will not consider the  $\sin(\phi_{S_A}+\phi_\pi^H)$
asymmetry in the sequel. A similar situation holds also for the
gluon generated $A_N^{\sin(\phi_{S_A}+2\phi_\pi^H)}$ asymmetry, where two
contributions analogous to the quark ones discussed above but for
linearly polarized gluons are involved.

\begin{figure*}[t]
\begin{center}
 \includegraphics[angle=0,width=0.4\textwidth]{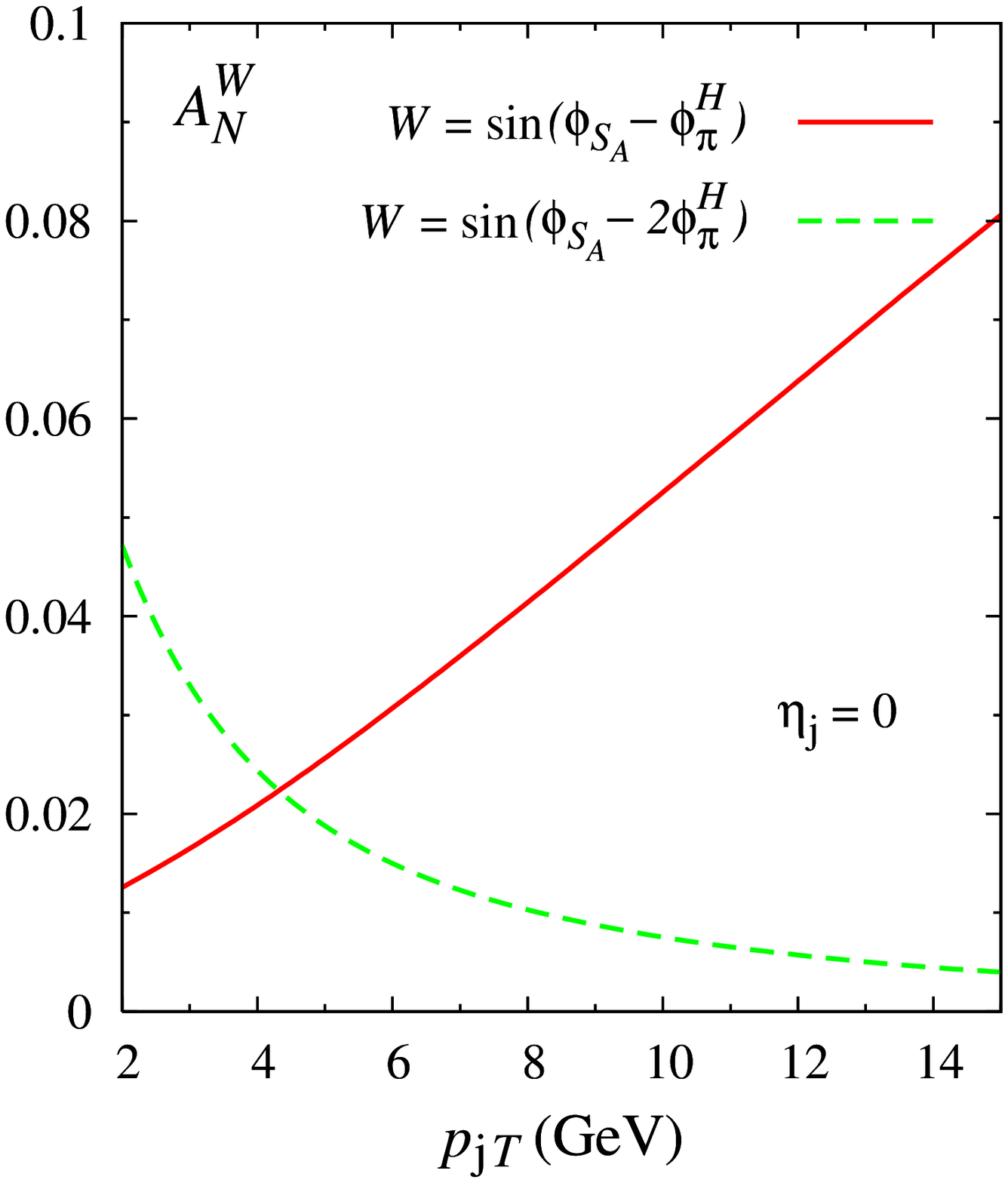}
 \includegraphics[angle=0,width=0.4\textwidth]{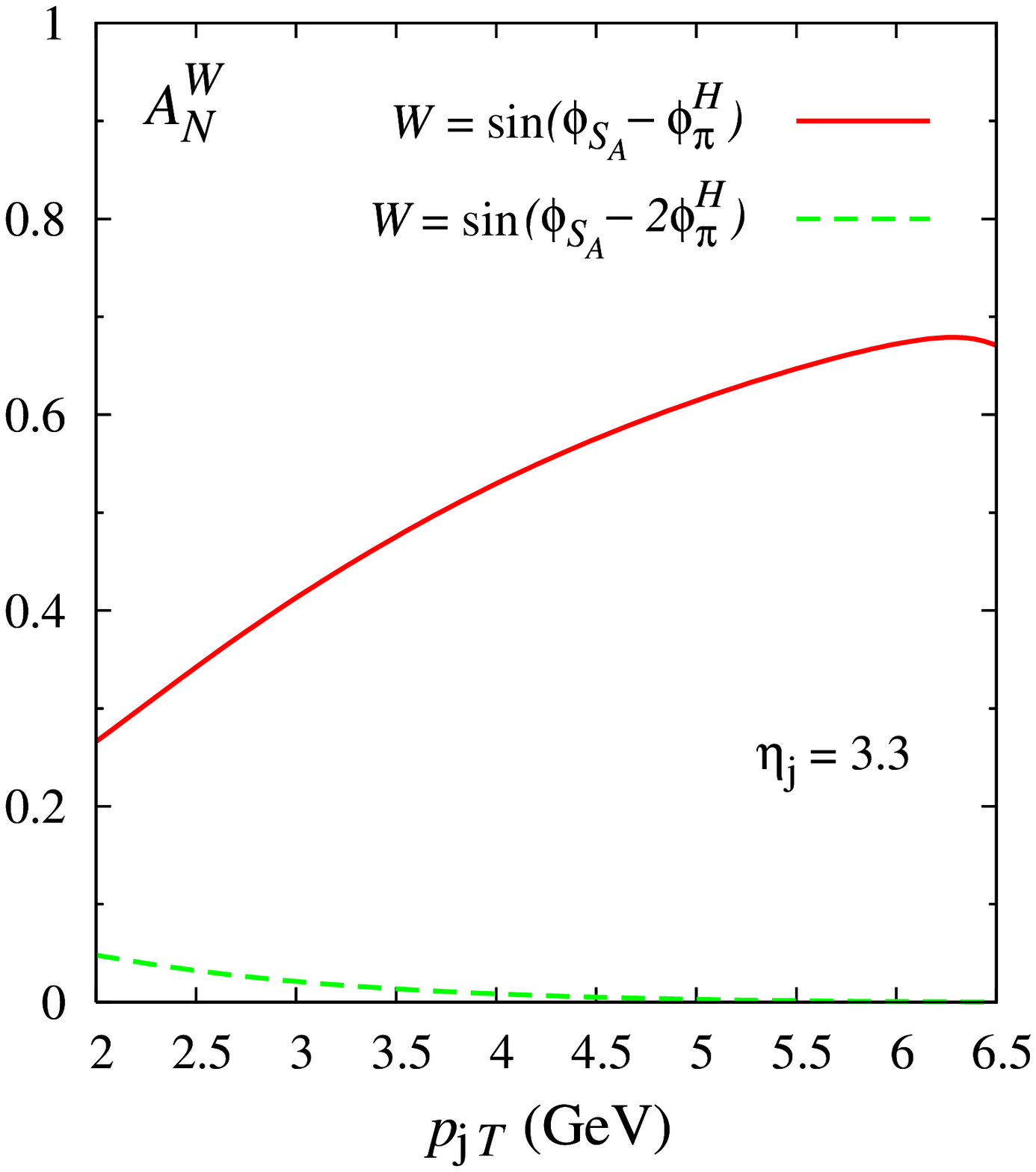}
 \caption{  Maximized  quark 
and gluon  Collins(-like) asymmetries
for the $p^\uparrow p\to {\rm jet}+\pi^+ + X$ process,
at $\sqrt{s}=200$ GeV  in two different rapidity regions
adopting
the Kretzer FF set. 
 \label{asy-an-coll-max200} }
\end{center}
\end{figure*}

\begin{figure*}[b]
\begin{center}
 \includegraphics[angle=0,width=0.4\textwidth]{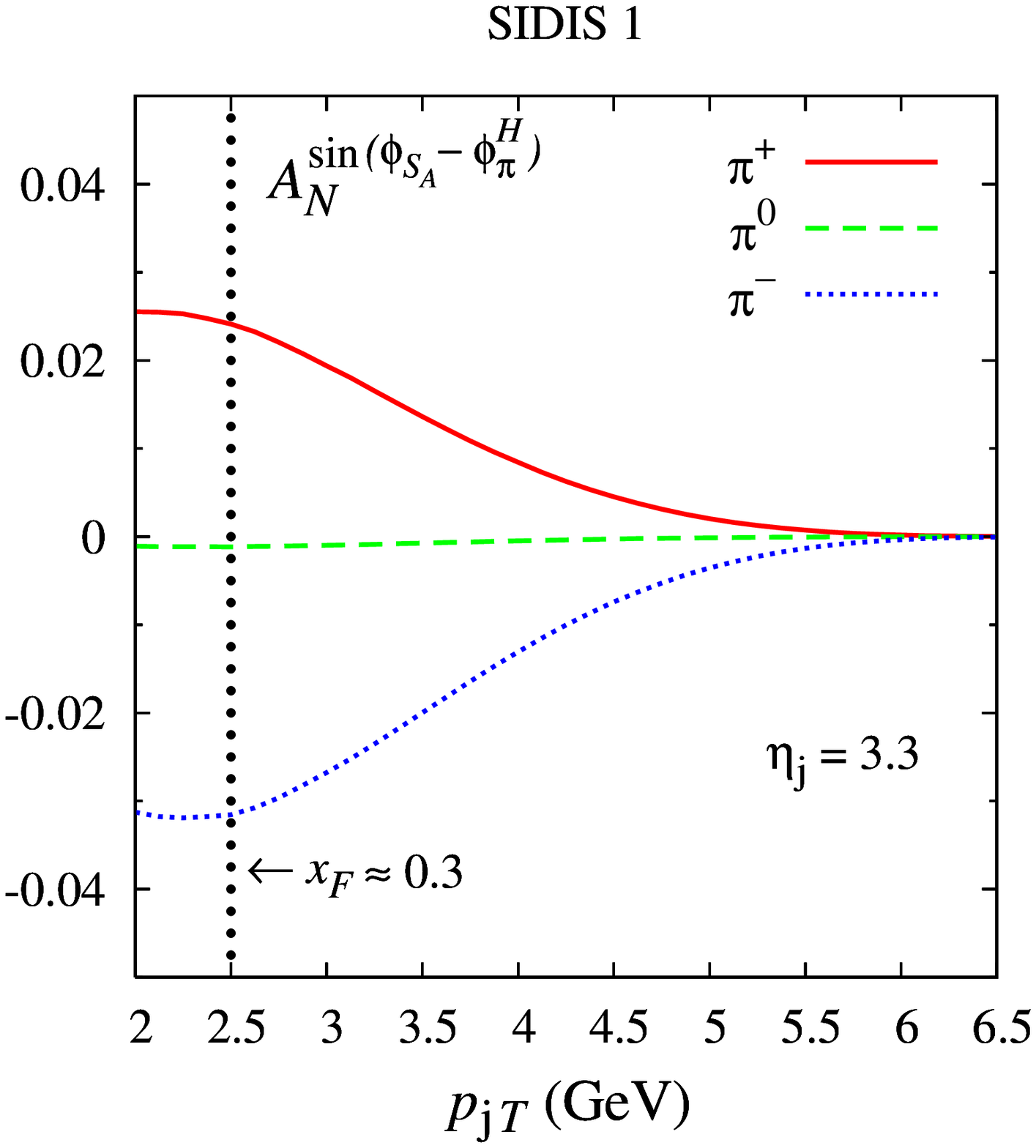}
 \includegraphics[angle=0,width=0.4\textwidth]{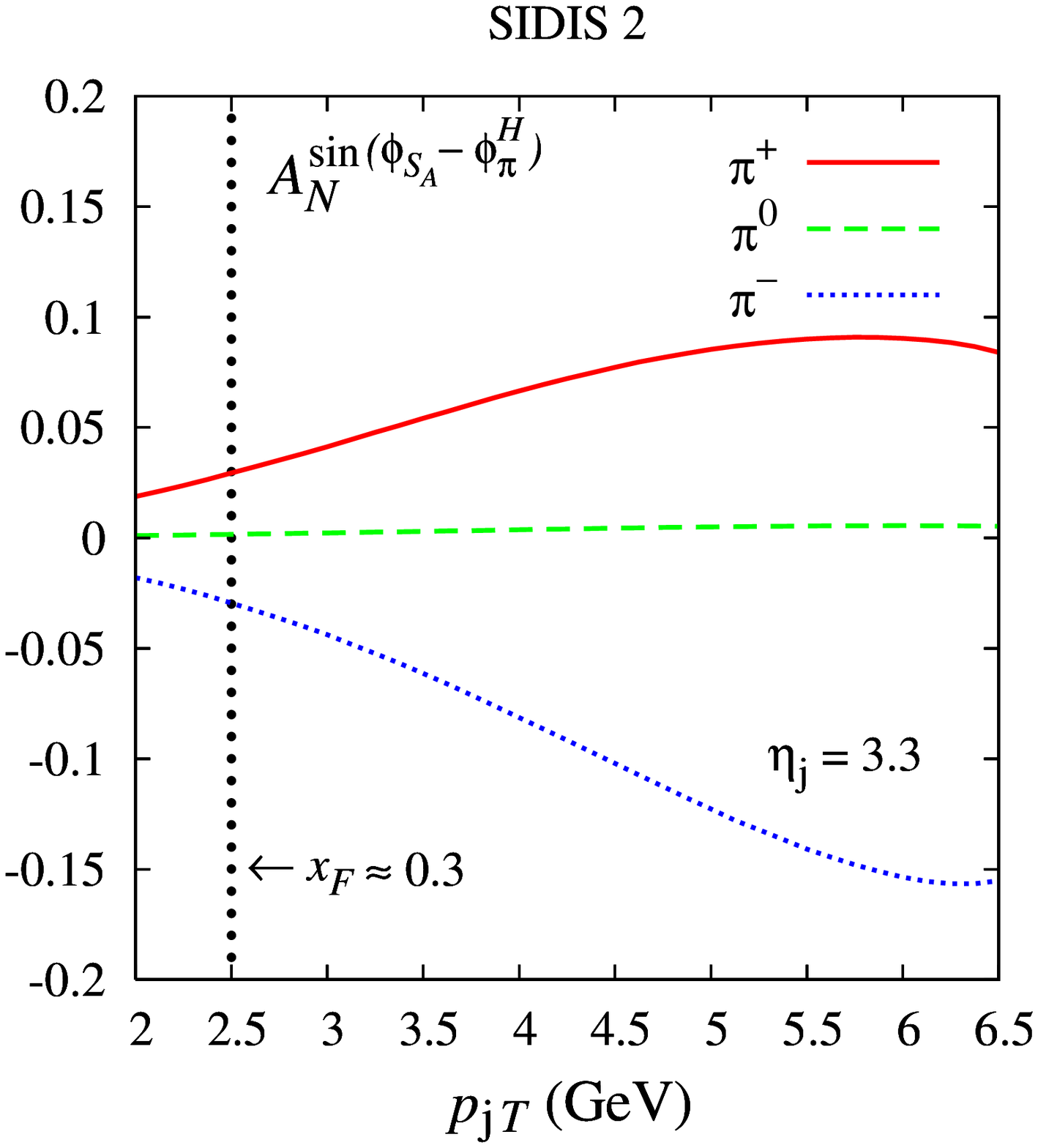}
 \caption{  The estimated quark Collins asymmetry
for the $p^\uparrow p\to {\rm jet}+\pi + X$ process,
obtained adopting the parameterizations SIDIS~1 
and SIDIS~2  respectively,
at $\sqrt{s}=200$ GeV  in the forward rapidity
region.
The dotted black vertical line delimits the region $x_F \approx 0.3$.
\label{asy-an-coll-par200} }
\end{center}
\end{figure*}

In Fig.~\ref{asy-an-coll-max200}
we present the quark $A_N^{\sin(\phi_{S_A}-\phi_\pi^H)}$ Collins asymmetry
 and the gluon $A_N^{\sin(\phi_{S_A}-2\phi_\pi^H)}$
Collins-like asymmetry  in the maximized scenario
in the central (left panel)
and forward (right panel) rapidity region as a function of $p_{{\rm j}\,T}$,
from $p_{{\rm j}\,T}=2$ GeV up to the maximum allowed value, adopting
the Kretzer FF set.
In the central rapidity region
the quark Collins asymmetry is very small at the lowest $p_{{\rm j}\,T}$
values, then increases almost linearly reaching about 8\% at the upper range.
Instead, in the forward rapidity region the asymmetry is (potentially)
always large and increases almost linearly from about 25\% to about 70\%
going from the lowest to the largest $p_{{\rm j}\,T}$ values.
Concerning the gluon Collins-like asymmetry, both in the central
and in the forward rapidity regions it is of the order of 5\%
at the lowest $p_{{\rm j}\,T}$ values, then starts decreasing slowly
and becomes negligible at large $p_{{\rm j}\,T}$ values.
Similar results hold 
when adopting the DSS set.

We consider now, for both neutral and charged pions,
numerical results for the quark Collins
asymmetry obtained adopting the parameterizations SIDIS~1 and
SIDIS~2 for the transversity distribution and the Collins
fragmentation function (no parameterizations are available yet
in the analogous gluon case).
It turns out that in the central rapidity region all the estimated
asymmetries are practically negligible.
Concerning the forward rapidity region, our results are shown
in Fig.~\ref{asy-an-coll-par200}.
The Collins asymmetry for neutral pions 
results to be almost vanishing.
For charged pions, similarly to the case of the Sivers asymmetry, the two
parameterizations give comparable results
only in the $p_{{\rm j}\,T}$
region where the transversity distribution is reasonably
constrained by SIDIS data (see the dotted black vertical line).
A measurement of this asymmetry
would be then very important and helpful in clarifying
the large $x$ behaviour of the quark transversity distribution.

\section{\label{sec-conclusions} Conclusions}
We have presented a study of the azimuthal asymmetries measurable
in the distribution of leading pions inside a large-$p_T$
jet produced in  unpolarized and single-transverse polarized proton proton
collisions for kinematical configurations accessible at RHIC.
To this end, we have adopted a generalized TMD parton model approach
with inclusion of spin and intrinsic parton motion effects both
in the distribution and in the fragmentation sectors.

In contrast to inclusive  pion 
production, where the Sivers and Collins mechanisms cannot be separated \cite{D'Alesio:2010am}, and  in close 
analogy with the SIDIS case, the leading-twist azimuthal asymmetries
 discussed above 
allow one
to discriminate among different effects by taking suitable moments of the asymmetries.  In principle, quark and gluon originating jets can also be distinguished, at least in some kinematical regimes. 
Hence, the proposed phenomenological analysis could be very helpful, 
for example,
 in clarifying 
the role played by the quark(gluon) Sivers distribution and by the Collins(-like) fragmentation function in the sizable single spin asymmetries observed
at RHIC for forward pion production. 
At the same time  it will give us the opportunity of testing the
factorization and universality assumptions, and of gaining
information on the size and \textit{sign} of the TMD functions discussed, in a kinematic region not covered by SIDIS data.

We finally stress that the unambiguous measurement of any of the
asymmetries, other than the Collins one, discussed above  would be a 
clear indication of the role played by intrinsic parton motion in the initial
 hadrons for the spin asymmetry sector in polarized hadronic 
collisions.

\section*{Acknowledgments}
C.P.~is supported by Regione Autonoma della Sardegna (RAS) through a research
grant under the PO Sardegna FSE 2007-2013, L.R. 7/2007, ``Promozione della
ricerca scientifica e dell'innovazione tecnologica in Sardegna".~U.D.~and F.M.~acknowledge partial support by Italian Ministero dell'Istruzione,
dell'Universit\`{a} e della Ricerca Scientifica (MIUR) under Cofinanziamento PRIN 2008,
and by the European Community under the FP7 ``Integrating Activities" project
``Study of Strongly Interacting Matter" (HadronPhysics2), grant agreement No.~227431.

\end{document}